\documentclass[manuscript]{aastex}

\newcommand{\reffig}[1]{Figure~\ref{#1}}  
\newcommand{\reftab}[1]{Table~\ref{#1}}   
\def\etal{et~al.}

\usepackage{lineno}
\usepackage{url}

\shorttitle{Observations of B2~1215+30 with VERITAS}
\shortauthors{VERITAS collaboration}

\begin{document}

\title{Long term observations of B2~1215+30 with VERITAS}

\author{
E.~Aliu\altaffilmark{1},
S.~Archambault\altaffilmark{2},
T.~Arlen\altaffilmark{3},
T.~Aune\altaffilmark{3},
M.~Beilicke\altaffilmark{4},
W.~Benbow\altaffilmark{5},
R.~Bird\altaffilmark{6},
A.~Bouvier\altaffilmark{7},
J.~H.~Buckley\altaffilmark{4},
V.~Bugaev\altaffilmark{4},
A.~Cesarini\altaffilmark{8},
L.~Ciupik\altaffilmark{9},
M.~P.~Connolly\altaffilmark{8},
W.~Cui\altaffilmark{10},
J.~Dumm\altaffilmark{11},
M.~Errando\altaffilmark{1},
A.~Falcone\altaffilmark{12},
S.~Federici\altaffilmark{13,14},
Q.~Feng\altaffilmark{10},
J.~P.~Finley\altaffilmark{10},
P.~Fortin\altaffilmark{5,15},
L.~Fortson\altaffilmark{11},
A.~Furniss\altaffilmark{7},
N.~Galante\altaffilmark{5},
L.~G\'{e}rard\altaffilmark{13},
G.~H.~Gillanders\altaffilmark{8},
S.~Griffin\altaffilmark{2},
J.~Grube\altaffilmark{9},
G.~Gyuk\altaffilmark{9},
D.~Hanna\altaffilmark{2},
J.~Holder\altaffilmark{16},
G.~Hughes\altaffilmark{13},
T.~B.~Humensky\altaffilmark{17},
P.~Kaaret\altaffilmark{18},
M.~Kertzman\altaffilmark{19},
Y.~Khassen\altaffilmark{6},
D.~Kieda\altaffilmark{20},
H.~Krawczynski\altaffilmark{4},
F.~Krennrich\altaffilmark{21},
M.~J.~Lang\altaffilmark{8},
A.~S~Madhavan\altaffilmark{21},
G.~Maier\altaffilmark{13},
P.~Majumdar\altaffilmark{3,22},
S.~McArthur\altaffilmark{23},
A.~McCann\altaffilmark{24},
P.~Moriarty\altaffilmark{25},
R.~Mukherjee\altaffilmark{1},
D.~Nieto\altaffilmark{17},
A.~O'Faol\'{a}in de Bhr\'{o}ithe\altaffilmark{6},
R.~A.~Ong\altaffilmark{3},
M.~Orr\altaffilmark{21},
A.~N.~Otte\altaffilmark{26},
N.~Park\altaffilmark{23},
J.~S.~Perkins\altaffilmark{27,28},
M.~Pohl\altaffilmark{14,13},
A.~Popkow\altaffilmark{3},
H.~Prokoph\altaffilmark{13},
J.~Quinn\altaffilmark{6},
K.~Ragan\altaffilmark{2},
L.~C.~Reyes\altaffilmark{29},
P.~T.~Reynolds\altaffilmark{30},
G.~T.~Richards\altaffilmark{26},
E.~Roache\altaffilmark{5},
D.~B.~Saxon\altaffilmark{16},
G.~H.~Sembroski\altaffilmark{10},
C.~Skole\altaffilmark{13},
A.~W.~Smith\altaffilmark{20},
M.~Soares-Furtado\altaffilmark{7},
D.~Staszak\altaffilmark{2},
I.~Telezhinsky\altaffilmark{14,13},
G.~Te\v{s}i\'{c}\altaffilmark{2},
M.~Theiling\altaffilmark{10},
A.~Varlotta\altaffilmark{10},
V.~V.~Vassiliev\altaffilmark{3},
S.~Vincent\altaffilmark{13},
S.~P.~Wakely\altaffilmark{23},
T.~C.~Weekes\altaffilmark{5},
A.~Weinstein\altaffilmark{21},
R.~Welsing\altaffilmark{13},
D.~A.~Williams\altaffilmark{7},
B.~Zitzer\altaffilmark{31}\\
(VERITAS collaboration)\\
\and
M.~B\"{o}ttcher\altaffilmark{32,33},
M.~Fumagalli\altaffilmark{34,35,36}, 
J.~Jadhav\altaffilmark{32}
}

\altaffiltext{1}{Department of Physics and Astronomy, Barnard College, Columbia University, NY 10027, USA}
\altaffiltext{2}{Physics Department, McGill University, Montreal, QC H3A 2T8, Canada}
\altaffiltext{3}{Department of Physics and Astronomy, University of California, Los Angeles, CA 90095, USA}
\altaffiltext{4}{Department of Physics, Washington University, St. Louis, MO 63130, USA}
\altaffiltext{5}{Fred Lawrence Whipple Observatory, Harvard-Smithsonian Center for Astrophysics, Amado, AZ 85645, USA}
\altaffiltext{6}{School of Physics, University College Dublin, Belfield, Dublin 4, Ireland}
\altaffiltext{7}{Santa Cruz Institute for Particle Physics and Department of Physics, University of California, Santa Cruz, CA 95064, USA}
\altaffiltext{8}{School of Physics, National University of Ireland Galway, University Road, Galway, Ireland}
\altaffiltext{9}{Astronomy Department, Adler Planetarium and Astronomy Museum, Chicago, IL 60605, USA}
\altaffiltext{10}{Department of Physics, Purdue University, West Lafayette, IN 47907, USA }
\altaffiltext{11}{School of Physics and Astronomy, University of Minnesota, Minneapolis, MN 55455, USA}
\altaffiltext{12}{Department of Astronomy and Astrophysics, 525 Davey Lab, Pennsylvania State University, University Park, PA 16802, USA}
\altaffiltext{13}{DESY, Platanenallee 6, 15738 Zeuthen, Germany}
\altaffiltext{14}{Institute of Physics and Astronomy, University of Potsdam, 14476 Potsdam-Golm, Germany}
\altaffiltext{15}{Laboratoire Leprince-Ringuet, Ecole Polytechnique, CNRS/IN2P3, F-91128 Palaiseau, France}
\altaffiltext{16}{Department of Physics and Astronomy and the Bartol Research Institute, University of Delaware, Newark, DE 19716, USA}
\altaffiltext{17}{Physics Department, Columbia University, New York, NY 10027, USA}
\altaffiltext{18}{Department of Physics and Astronomy, University of Iowa, Van Allen Hall, Iowa City, IA 52242, USA}
\altaffiltext{19}{Department of Physics and Astronomy, DePauw University, Greencastle, IN 46135-0037, USA}
\altaffiltext{20}{Department of Physics and Astronomy, University of Utah, Salt Lake City, UT 84112, USA}
\altaffiltext{21}{Department of Physics and Astronomy, Iowa State University, Ames, IA 50011, USA}
\altaffiltext{22}{Saha Institute of Nuclear Physics, Kolkata 700064, India}
\altaffiltext{23}{Enrico Fermi Institute, University of Chicago, Chicago, IL 60637, USA}
\altaffiltext{24}{Kavli Institute for Cosmological Physics, University of Chicago, Chicago, IL 60637, USA}
\altaffiltext{25}{Department of Life and Physical Sciences, Galway-Mayo Institute of Technology, Dublin Road, Galway, Ireland}
\altaffiltext{26}{School of Physics and Center for Relativistic Astrophysics, Georgia Institute of Technology, 837 State Street NW, Atlanta, GA 30332-0430}
\altaffiltext{27}{CRESST and Astroparticle Physics Laboratory NASA/GSFC, Greenbelt, MD 20771, USA.}
\altaffiltext{28}{University of Maryland, Baltimore County, 1000 Hilltop Circle, Baltimore, MD 21250, USA.}
\altaffiltext{29}{Physics Department, California Polytechnic State University, San Luis Obispo, CA 94307, USA}
\altaffiltext{30}{Department of Applied Physics and Instrumentation, Cork Institute of Technology, Bishopstown, Cork, Ireland}
\altaffiltext{31}{Argonne National Laboratory, 9700 S. Cass Avenue, Argonne, IL 60439, USA}


\altaffiltext{32}{Centre for Space Research, North-West University,
  Potchefstroom Campus, Potchefstroom 2520, South Africa}
\altaffiltext{33}{Astrophysical Institute, Department of Physics and
  Astronomy, Ohio University, Athens, OH 45701, USA}
\altaffiltext{34}{Carnegie Observatories, 813 Santa Barbara Street,
  CA-91101 Pasadena, USA} 
\altaffiltext{35}{Department of Astrophysics, Princeton University,
  NJ-08544-1001 Princeton, USA} 
\altaffiltext{36}{Hubble Fellow}


\begin{abstract}
We report on VERITAS observations of the BL Lac object B2~1215+30
between 2008 and 2012. During this period, the source was detected at
very high energies (VHE; E$>$100~GeV) by VERITAS with a significance
of $8.9\sigma$ and showed clear variability on time scales larger than
months.  
In 2011, the source was found to be in a relatively bright state and
a power-law fit to the differential photon spectrum yields a spectral
index of $3.6 \pm 0.4_{\mathrm{stat}} \pm 0.3_{\mathrm{syst}}$ with an
integral flux above 200~GeV of 
$(8.0 \pm 0.9_{\mathrm{stat}} \pm 3.2_{\mathrm{syst}}) 
\times 10^{-12}\, \mathrm{cm}^{-2} \mathrm{s}^{-1}$. 
No short term variability could be detected during the bright state in
2011. 
Multi-wavelength data were obtained contemporaneous with the VERITAS
observations in 2011 and cover optical (Super-LOTIS, MDM, 
\textit{Swift}-UVOT), X-ray (\textit{Swift}-XRT), and gamma-ray 
(\textit{Fermi}-LAT) frequencies. These were used to construct the
spectral energy distribution (SED) of B2~1215+30. 
A one-zone leptonic model is used to model the blazar emission and the
results are compared to those of MAGIC from early 2011 and other
VERITAS-detected blazars. The SED can be well reproduced
with model parameters typical for VHE-detected BL Lacs. 
\end{abstract}



\section{Introduction}
B2~1215+30, also commonly referred to as ON~325 or 1ES~1215+303, was
first detected in the Bologna Northern Cross telescope survey
conducted at 408 MHz \citep{1970A&AS....1..281C}. It was one of the
first BL Lac-type objects to be identified \citep{1971Natur.231..515B}
and was one member of the small set of objects used to define the
class. The distance to this source is uncertain and two different
redshift values, both obtained from spectroscopic measurements, can be
found in the literature: 
z~=~0.130 (\citealp{2003ApJS..148..275A};
NED\footnote{\url{http://ned.ipac.caltech.edu/}}) 
and z~=~0.237 (\citealp{1993ApJS...84..109L}; 
Simbad\footnote{\url{http://simbad.u-strasbg.fr/simbad/}}). 
A 10-minute exposure with the FAST instrument on the FLWO 60''
telescope in 2011 did not yield any obvious emission lines in the
continuum spectrum to resolve this discrepancy (E.~Falco, priv. comm.).  
Similarly, no spectral features were evident in a high SN 
spectrum (SN 50-120 from the red to blue side) we obtained with the Lick
Observatory Kast double spectrograph on the Shane 3-m telescope on 13
February 2013 (MJD 56336). 

%
%
BL Lac objects and flat spectrum radio quasars (FSRQs)
belong to the most extreme sub-class of active galactic nuclei (AGN),
named blazars. Their relativistic jet is oriented close to the
observer's line-of-sight. They show rapid variability at all
wavelengths with the fastest being observed at very high energies
(VHE; E$>$100~GeV) on time scales of minutes 
\citep{2007ApJ...664L..71A,2011ApJ...730L...8A,2013ApJ...762...92A}. 
%
%
The spectral energy distribution (SED) of blazars is dominated by
non-thermal emission and consists of two distinct, broad components. 
The low-energy component ranges from radio to UV/X-rays and is
widely believed to be due to synchrotron emission from
ultra-relativistic electrons in the jet magnetic field. 
To explain the origin of the second component, peaking between X-rays and
gamma rays, two fundamentally different scenarios exist, dominated by 
either leptonic or hadronic emission. 
In leptonic scenarios, the high-energy radiation is produced via
inverse-Compton scattering of the ultra-relativistic electrons
responsible for 
the synchrotron emission. Possible seed photons are synchrotron
photons within the jet (synchrotron self-Compton, SSC model), or
external photons (external Compton, EC model) from the disk, the
broad line region, or the jet. 
In hadronic scenarios, protons are accelerated to sufficiently high
energies and the high-energy emission is dominated by neutral pion
decay as well as hadronic synchrotron radiation.  
For a review of different blazar models see 
\citet{2012arXiv1205.0539B} and references therein.

Based on its SED, B2 1215+30 was suggested 
by \citet{2002A&A...384...56C} as a potential TeV source. It is
now classified as a bright intermediate-frequency-peaked BL Lac object
(IBL) based on its 
synchrotron peak location at $10^{15.6}$~Hz \citep{2006A&A...445..441N}. 
It is listed in the \textit{Fermi} bright-source list
\citep{2009ApJS..183...46A}, and  appears in later \textit{Fermi} catalogs 
\citep[e.g.][]{2011ApJ...743..171A}, where it is classified as a
high-synchrotron-peaked BL Lac (HSP).

In early January 2011, B2~1215+30 was detected in the VHE band by
MAGIC during observations triggered by an optical high state
\citep{2011ATel.3100....1M}. 
The flux above 200~GeV was $(7.7 \pm 0.9) \times 10^{-12} \,
\mathrm{cm}^{-2}\, \mathrm{s}^{-1}$ with a photon spectral index of
$\Gamma = 2.96 \pm 0.14$ \citep{2012A&A...544A.142A}.

In this paper we report on the results of VERITAS observations
taken in the direction of B2~1215+30 between December 2008 and
May 2012. This blazar is in the same field of view as
1ES~1218+304\footnote{The two sources are 0.76\degr\ away from each other.}, a
bright VHE blazar which is regularly observed 
by VERITAS \citep{2011arXiv1110.0038W}. 
A large part of the data presented here originates from observations
taken on 1ES~1218+304. 


\section{VERITAS: VHE gamma-ray observations}
%
%
VERITAS is an array of four imaging atmospheric Cherenkov telescopes 
located in southern Arizona. It is sensitive to gamma-ray energies
from 100~GeV to about 30~TeV and has been fully operational since Fall 2007. 
Short Cherenkov light flashes produced in extensive air showers are
focused by 12~m diameter reflectors onto fast-recording
cameras. Each camera is equipped with 499 photomultiplier tubes with a
total field of view of 3.5\degr. 
In Summer 2009, one of the four telescopes was moved to a new
location. This yielded about $30\%$
sensitivity improvement and reduced the observation time needed to
detect a 1\% Crab Nebula-like source with 5 standard deviations
($5\sigma$) from 48~hours to less than 30~hours \citep{2011arXiv1111.1225H}. 

%
%
The observations reported here include observations of B2~1215+30 and
1ES~1218+304, two sources which are separated by 0.76\degr. All
observations were taken in ``wobble mode'' where the source
position is offset by 0.5\degr\ from the camera center to
allow for simultaneous background estimation 
\citep[e.g.][]{2007A&A...466.1219B}. Combining the 
observations on both sources, VERITAS observed B2~1215+30 for more
than 93 hours between December 2008 and May 2012. 
The data are divided into three data sets, corresponding to yearly
observation epochs. The first one spans 34 hours from December 2008 to
May 2009 at a mean zenith angle of 20\degr, the second data set was
recorded between January and June 2011 (42 hours) at a mean zenith
angle of 15\degr, and the third data set was taken from January to
May 2012 (17.5 hours) with a mean zenith angle of 12\degr. 
Most of the observations presented here had 
1ES~1218+304 as the principal target, resulting in different pointing
offsets from the position of B2~1215+30 (from 0.3\degr\ to 1.3\degr). 
Given that the radial acceptance of the camera is not flat, this causes a
lower average sensitivity for the VERITAS exposure. Correcting 
for this effect, the total effective exposures on B2~1215+30, are 29,
38, and 15 hours for the different observation epochs. 

%
%
After run selection and nightly calibration, image cleaning is 
performed to remove the night sky background contamination from the
shower images. These images are then parameterized using a
second-moment analysis \citep{1985ICRC....3..445H}. Additionally, a  
log-likelihood fitting algorithm is applied to recover truncated 
images at the edge of the camera. After image quality cuts, 
the shower direction and core location are reconstructed for events having a
minimum of three telescope images. 
For each event the energy is estimated from lookup tables, with 
an energy resolution of about 15-20\%. 
To separate the gamma-ray events from the hadronic events, a set of optimized
cuts based on image parameters is applied, as described in
\citet{2008ApJ...679.1427A}. The optimization of those cuts has been
performed \textit{a priori} on a 5\% Crab Nebula-like source and yielded an
energy threshold of about 250~GeV for observations at 20\degr\ zenith
angle. 
The remaining background is estimated using a ring-background
model. The ON region is circular,
centered on the source position 
with radius $\theta \leq 0.09\degr$. The OFF region is defined as a
ring, placed around the ON region, with inner and outer radii of
0.46\degr\ and 0.54\degr, respectively. The radii are chosen so that
the ratio between ON and OFF area is 1:10. The normalization $\alpha$
is given by the area ratio modified by the radial acceptance of the camera. 
Regions around bright stars (V magnitude $<$~7), as well as the region
around the position of 1ES~1218+304, were excluded from the background
estimation.  

%
%
The analysis of the total data set over the time period from December
2008 to May 2012 yields 259 events excess over background. 
The resulting detection significance is $8.9 \sigma$ according to
Eq.~17 in \citet{1983ApJ...272..317L}. 
The results of the three observing periods are presented in
\reftab{tbl-1}. 
In 2011, the source was clearly detected with a significance of 
$10.4 \sigma$, while in 2008/09 and in 2012 the source is not detected
with a significance greater than $5 \sigma$. 
Before presenting the results of these two latter periods, we
concentrate on the 2011 data set where the significant detection
allows for a spectral analysis.

\begin{figure}
\epsscale{0.5} 
\plotone{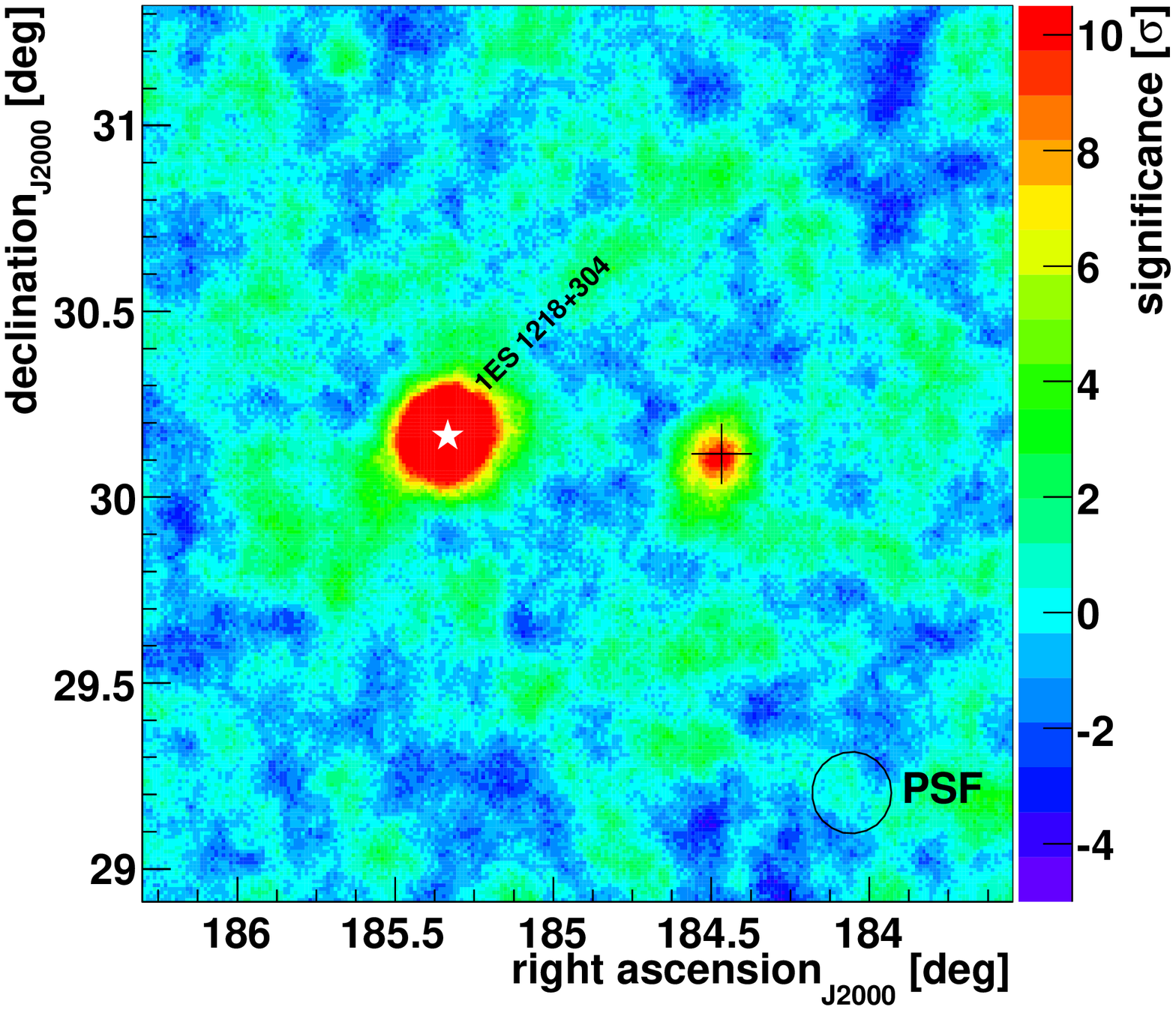}
\caption{VERITAS significance sky map of 2011. The black cross shows the
  position of B2~1215+30 and the white star indicates the position of
  1ES~1218+304. Both sources are point-like, but appear to have 
  a different size due to the saturation of the color
  scale. \label{fig1}} 
\end{figure}

%
%
\reffig{fig1} shows the significance sky map for the 2011
data set. 
To determine the position of the VHE gamma-ray emission, a symmetric
2D Gaussian was fitted to the excess sky map (binned to
$0.05\degr$) of this data set. It revealed a point-like excess with
the best-fit source position at R.A. = $12^h17^m48.5^s \pm 1.7^s$, 
DEC = $+30\degr06'06'' \pm 25''$ with a systematic uncertainty of $50''$. 
The VERITAS source is thus named VER~J1217+301, and is positionally
consistent with the BL Lac object B2~1215+30 \citep{1998AJ....116..516M}.

\begin{figure}
\epsscale{0.5} 
\plotone{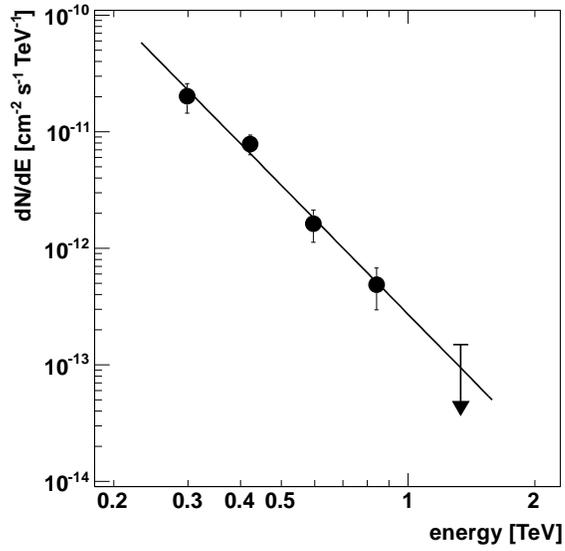}
\caption{Differential photon spectrum of B2~1215+30 obtained with
  VERITAS in 2011. The spectral points are fitted with a
  power-law. The error bars denote $1\sigma$ uncertainties, and an
  95\% upper limit is drawn for a spectral point with less then
  $2 \sigma$ significance.\label{fig2}}
\end{figure}

\begin{figure}
\epsscale{0.7} 
\plotone{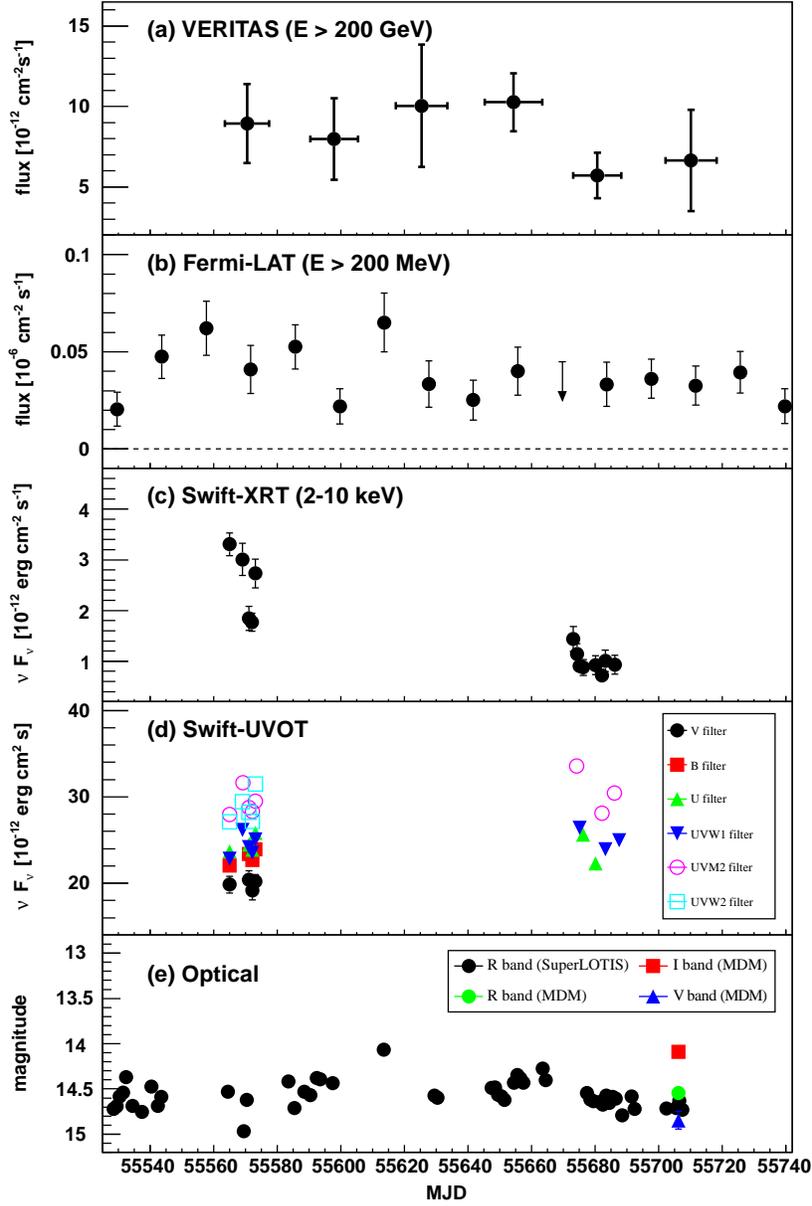}
\caption{Multi-wavelength light curve of B2~1215+30 for the first
  half of 2011. (a) Monthly binned VHE gamma-ray light curve above
  200~GeV as measured by VERITAS. (b) Bi-weekly binned
  {\it Fermi}-LAT light curve above 200~MeV. (c) X-ray light
  curve measured by {\it Swift}-XRT. (d) {\it Swift}-UVOT light
  curve for the different filters (given in the legend). 
  (e) Optical light curve. The black points are measured by
  Super-LOTIS in the R-band, with a statistical error of $\sim
  0.1-0.2$ mag. The V, R, and I points are from MDM
  observations. \label{fig3}} 
\end{figure}

%
%
The derived differential photon spectrum of the 2011 data set is shown
in \reffig{fig2}. It can be fitted by a power law ($\chi^2$/ndf = 1.25/2): 
dN/dE = $F_0 (E/300\,\mathrm{GeV})^{-\Gamma}$, 
with $F_0 = (2.3 \pm 0.5_{\mathrm{stat}} \pm 0.9_{\mathrm{syst}})
\times 10^{-11}$ cm$^{-2}$ s$^{-1}$ 
and $\Gamma = 3.6 \pm 0.4_{\mathrm{stat}} \pm 0.3_{\mathrm{syst}}$. 
The flux above 200~GeV is $(8.0 \pm 0.9_{\mathrm{stat}} \pm 3.2_{\mathrm{syst}}) 
\times 10^{-12}\, \mathrm{cm}^{-2} \mathrm{s}^{-1}$. 
This corresponds to 3.4\% of the Crab Nebula flux
\citep{1998ApJ...503..744H} above the same energy threshold. 
A 29.5-day binned light curve above 200~GeV is produced and shown in 
\reffig{fig3}. A constant fit to these flux points showed no evidence
for deviation from a steady flux ($\chi^2/ndf = 4.7/5$). 
No significant flux variations within any monthly bin were detected
either.

%
%
An analysis of the data available outside the 2011 season revealed
lower gamma-ray fluxes (see \reftab{tbl-1}). The 2008/09 data set
analysis resulted in a gamma-ray excess of $1.1\sigma$ significance at
the source location. 
Using the method of \citet{1983NIMPR.212..319H}, this excess
corresponds to a 99\% upper limit above 200~GeV of 2\% of the Crab
Nebula flux, assuming the same spectral index as derived in 2011. 
In 2012, the source is observed with a significance of
$3.2\sigma$. Given that it is an established VHE emitter a flux is
derived: 
$F(E>200\mathrm{GeV}) = (2.8 \pm 1.1_{\mathrm{stat}} \pm 1.1_{\mathrm{syst}}) 
\times 10^{-12}\, \mathrm{cm}^{-2} \mathrm{s}^{-1}$,  
corresponding to about 1.2\% of the Crab Nebula flux above the same
energy threshold. 
The hypothesis of a constant flux between the three seasons is
excluded at the level of $4.5\sigma$. 
This shows that the source was significantly fainter in 2008/09 and 2012
compared to the relatively bright flux state in 2011, as shown both by
MAGIC and VERITAS measurements.


\section{Fermi-LAT: High-energy gamma-ray observations}
The Large Area Telescope (LAT) on board the {\it Fermi} satellite
is a pair-conversion gamma-ray telescope sensitive to photon 
energies from 20 MeV to a few hundred GeV
\citep{2009ApJ...697.1071A}. 
A binned likelihood  analysis was performed using the LAT ScienceTools
(version v9r23p1) and P7SOURCE\_V6 instrument response functions. 
``Diffuse'' class events with $0.2 < E/\mathrm{GeV} < 100$ in a
square region of interest (ROI) of $20\degr \times 20\degr$ around
B2~1215+30 were selected. 
The center of the ROI was shifted by 4\degr\ towards the bright FSRQ
4C~+21.35 (8.87\degr\ away from B2~1215+30) to avoid edge effects. Further quality selection was performed by rejecting events with a
zenith angle $> 100\degr$ and a rocking angle $> 52\degr$ in order to
avoid contamination from albedo photons from the Earth's limb.
A background model was constructed including nearby gamma-ray sources
and diffuse emission. All known gamma-ray sources from the 
second \textit{Fermi} catalogue \cite[2FGL; ][]{2012ApJS..199...31N} within
the ROI were included in the model. Sources outside the
ROI, but within 5\degr\ of the ROI edges, were also included to
account for possible photon contamination due to the large LAT point
spread function. As in the 2FGL catalogue, a log-parabola function was
used for sources with significant spectral curvature. Otherwise,
spectra were described as a power law. The spectral parameters of the
sources inside the ROI were left free during the fitting
procedure. Sources outside the ROI, but within 5\degr\ of the
ROI edges, had their spectral parameters fixed to the 2FGL catalog values. 
The galactic and extragalactic diffuse gamma-ray emission together 
with the residual instrumental background were also modeled using the
publicly-available files\footnote{The files used were gal\_2yearp7v6\_v0.fits
  for the Galactic diffuse and iso\_p7v6source.txt for the isotropic
  diffuse component as available at 
  \url{http://fermi.gsfc.nasa.gov/ssc/data/access/lat/BackgroundModels.html}.}.

%
%
A 14-day binned light curve using the first 48 months of the {\it
Fermi} mission was produced. 
During the period quasi-simultaneous with the 2011 VERITAS observations
(MJD $55560-55720$), the flux above 200~MeV is compatible with being
constant ($\chi^2/ndf = 11.1/10$), and the light curve is shown in
\reffig{fig3}. 
%
%
A spectrum in the {\it Fermi}-LAT energy range was derived using this subset
of observations only. During that 160-day period, B2~1215+30
is detected with a test statistic value of $TS = 363$, corresponding
to a significance of about $19\sigma$. 
The spectrum is compatible with a power law with a photon index 
$\Gamma = 1.97 \pm 0.08$. The integral flux above 200~MeV is 
$(3.45 \pm 0.34) \times 10^{-8}\, \mathrm{cm}^{-2}\, \mathrm{s}^{-1}$. 
Potential contamination 
from the nearby source 1ES~1218+304 (at 0.76\degr\ distance) was
checked by producing a residual $TS$ map; no features or
asymmetries in the $TS$ distribution of B2~1215+30 were seen. 
It is worth noting that in the GeV range the flux of 1ES~1218+304 is
$\sim 0.4-0.6$ times that of B2~1215+30, in contrast to the VHE regime,
where 1ES~1218+304 is typically much brighter. 

Given the clear variability seen in the VHE band, a mean flux above
200~MeV contemporaneous with the 2008/09 and 2012 VERITAS observations
was derived and is 
$(1.8 \pm 0.3) \times 10^{-8}\, \mathrm{cm}^{-2}\, \mathrm{s}^{-1}$ and  
$(3.0 \pm 0.4) \times 10^{-8}\, \mathrm{cm}^{-2}\, \mathrm{s}^{-1}$,
respectively. The hypothesis of a constant flux in the high-energy
regime between the three seasons contemporaneous with the VERITAS
observations can be rejected at the $3\sigma$ level.


\section{Swift-XRT: X-ray observations}
The X-ray telescope (XRT) on board the {\it Swift} satellite is designed
to measure X-rays in the $0.2-10$ keV energy range
\citep{2005SSRv..120..165B}. Target of opportunity observations were 
obtained in January 2011 (MJD $55565-55573$), following the detection
of VHE emission from B2~1215+30, as well as on 10 nights in
April/May 2011 (MJD $555673-55686$). 
All XRT data presented here were taken in photon counting mode with
negligible pile-up effects. 
The data reduction and calibration were done using HEASoft, XSPEC
version 12.6.0 and the swxpc0to12s6\_20070901v011.rmf response
function. 
The data were grouped, requiring a minimum of 20 counts/bin, and then
fitted with an absorbed power law model. 
The galactic column density of $N_{\mathrm{H}} = 1.74 \times 10^{20}
\mathrm{cm}^{-2}$ was used, taken from the LAB neutral hydrogen survey
\citep{2005yCat.8076....0K}. 
When it was left free during the fit, the column density value was
consistent with what was found by the LAB survey.

The spectral analysis of the two time periods shows the blazar in different
states. The observations performed in January indicate a harder
and brighter flux state, allowing the data to be fitted with an absorbed
power law up to 10~keV. The highest integrated flux was found on MJD
55565 with $F_{[2-10\mathrm{keV}]} = (3.31 \pm 0.22) \,\times 10^{-12}
\,\mathrm{erg}\,\mathrm{cm}^{-2}\,\mathrm{s}^{-1}$ and 
a photon index of $2.46 \pm 0.05$. It will be referred to as the high
X-ray state in the SED modeling section. 
The observations taken in April/May show the object in a lower-flux state,
with too poor statistics in the individual observations in the energy
bins above 5~keV to constrain a 
spectral fit. However, combining the exposures from all the observations of
April/May allows a fit in the 0.4 to 10 keV range with an integrated
flux of $F_{[0.4-10\mathrm{keV}]} = (4.25 \pm 0.16)\, \times 10^{-12}
\,\mathrm{erg}\,\mathrm{cm}^{-2}\,\mathrm{s}^{-1}$ and 
a photon index of $2.74 \pm 0.04$. 
This average spectrum is then used to represent the low X-ray
state of the SED for the modeling in Section~6.

\begin{figure}
\epsscale{0.5} 
\plotone{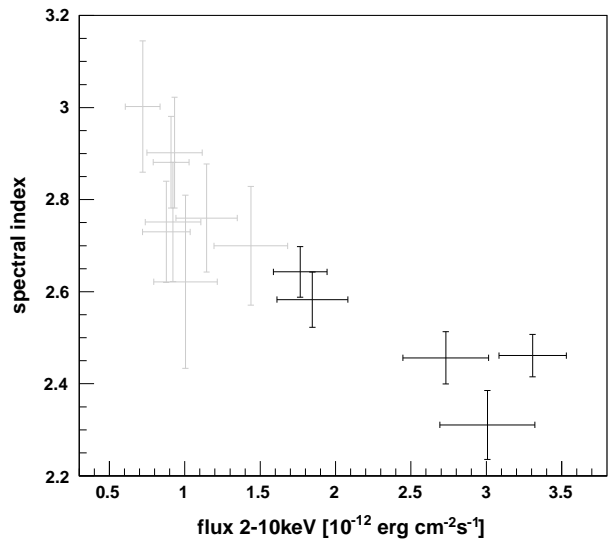}
\caption{Flux-spectral index correlation of {\it Swift}-XRT; 
  black points corresponding to the January observations and grey
  points to the April/May observations. The
  correlation coefficient was found to be $r=-0.88$, implying a
  strong (negative) correlation between integrated flux and spectral
  index. \label{fig4}}
\end{figure}

Additionally, an X-ray flux-index correlation study was performed on the
entire data set. The results are shown in
\reffig{fig4}. The correlation coefficient is found to
be $r = -0.88$ with an uncertainty of $< 0.1$. This implies a strong
(negative) correlation between spectral index and integrated flux of
the X-ray observations. 
A similar anti-correlation could also be seen in several other
VHE-emitting blazars, e.g., the IBLs VER~J0521+211 \citep{verj0521} or
W~Comae \citep{2009ApJ...707..612A}. However, this trend of ``harder
when brighter'' is not always been observed in VHE blazars, e.g. no
correlation between X-ray flux and its spectral index could be
detected during the low VHE-flux state of the blazar 1ES~1959+650
\citep{2013ApJ...775....3A}.


\section{UV and optical observations}

A {\it Swift} Ultra Violet and Optical Telescope 
\citep[UVOT;][]{2005SSRv..120...95R} analysis has been performed including all
the observations performed between January and May 2011. 
Exposures were taken in V, B, U, UVW1, UVM2 and UVW2 pass bands in
{\it image mode}, discarding the photon-timing information. 
The photometry was computed following
the general prescription of \citet{2008MNRAS.383..627P} and
\citet{2010MNRAS.406.1687B}, carefully excluding the
contribution from nearby faint objects.

A dedicated inter-calibration study between optical, UV and X-ray
datasets was carried out, adopting the $N_{\mathrm{H}}$ parameter for the hydrogen
column (obtained from the LAB survey). 
The results were reddening corrected using $E(B-V)=0.023$ mag 
\citep{1998ApJ...500..525S}. 
Then, the corresponding optical/UV galactic extinction coefficients
were computed ($R_V=2.667$) and applied
\citep{1999PASP..111...63F}. The host galaxy 
contribution of B2 1215+30 was estimated using the PEGASE-HR code
\citep{2004A&A...425..881L} extended for the ultraviolet UVOT filters
and by using the R-band photometric results of
\citet{2007A&A...475..199N}. 
No zodiacal light correction was introduced.
For each filter, the integrated flux was computed by using the
related effective frequency and not convolving the filter transmission
with the source spectrum. This may produce a moderate overestimation
(around $10$\%) of the integrated flux. The total upper limit
systematic uncertainty is 15\%.

%
%
In the optical regime, we monitored B2~1215+30 using the Super-LOTIS 
(Livermore Optical Transient Imaging System) robotic telescope over
the period December 2010 $-$ March 2012. 
In addition to these R-band observations, B2~1215+30 was observed with
the 1.3~m McGraw-Hill telescope of the MDM observatory, located at
Kitt Peak, Arizona, during one week in May 2011 (MJD $55706 - 55709$),
using standard V, R, and I filters. 
The data were bias-subtracted and flat-fielded using the
routines of the Image Reduction and Analysis Facitily (IRAF;
\citealp{1993ASPC...52..173T,1986SPIE..627..733T}).  
Comparative photometry with stars of known magnitude was performed 
and the resulting light curve is shown in \reffig{fig3} showing clear 
variability contemporanous with the VERITAS measurements. 
This is in line with the variability seen on the publicly available
light curves from the Tuorla 
Observatory\footnote{ \url{http://users.utu.fi/kani/1m/ON_325.html}}. 

For the construction of the optical SED using MDM observations, 
the magnitudes were corrected for Galactic extinction according to
\cite{1998ApJ...500..525S}.  
The values are A$_\mathrm{V}$ = 0.079, A$_\mathrm{R}$ = 0.064, and 
A$_\mathrm{I}$ = 0.046, as provided by the NED.


\section{Spectral energy distribution and modeling}

An SED was constructed using the multi-wavelength data obtained in
2011. During this time, no variability was detected in the high- 
or very-high-energy regimes. Given the low statistics in those energy
regimes, the \textit{Fermi}-LAT data contemporaneous with the 
VERITAS observations in 2011 are used (MJD $55560-55720$). 
Variability is clearly seen in X-rays and, 
therefore, two spectra are extracted: 
one to represent the high X-ray state in January (MJD
55565) and the other one to represent the low X-ray state observed in
April/May (using the combined spectrum from all observations between
MJD $55673-55686$). 
\textit{Swift}-UVOT data simultaneous with the X-ray
observations were used when available. 
Given the relatively large systematic uncertainty of the UVOT analysis,
the quasi-simultaneous optical spectrum from MDM (MJD $55706-55709$) is
additionally used in the SED representing the low X-ray state. 
To complete the low energy part of the SED, archival data in the
micrometer wavelength regime, taken from \citet{2004MNRAS.352..673A},
are included. Unfortunately, no information on the 
variability at those wavelengths is found, but since blazars are
usually variable at all wavelengths, the inclusion of these 
archival data in the modeling will be discussed later.

\begin{figure*}
\epsscale{0.8} 
\plotone{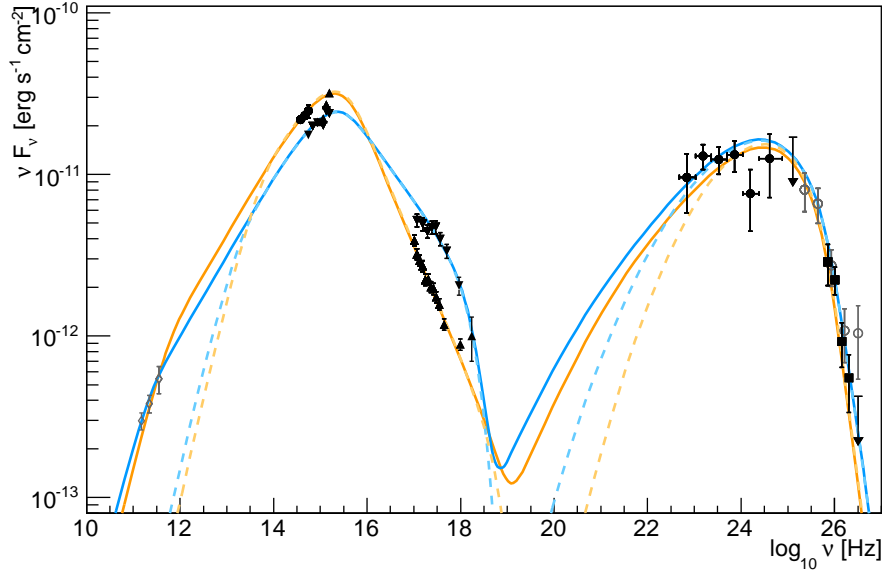}
\caption[Spectral energy distribution of B2~1215+30.]{Spectral
  energy distribution of B2~1215+30 during 2011 
  together with the SSC model for a redshift of $z=0.130$. 
  The data points are the following (from low to high
  frequencies):
  ($\Diamond$) archival data, 
  ($\bullet$) optical data from MDM,
  ($\blacktriangledown$) high X-ray state observed by
  \textit{Swift}-XRT \& UVOT, 
  ($\blacktriangle$) low X-ray state observed by
  \textit{Swift}-XRT \& UVOT, 
  ($\bullet$) \textit{Fermi}-LAT, 
  ($\blacksquare$) VERITAS, and 
  ($\circ$) MAGIC. 
  The blue lines represent the model for the high X-ray state data set
  and the orange lines the model for the low X-ray state, respectively. 
  Solid lines take the archival data in the micrometer regime 
  into account; dashed lines are the model using contemporaneous
  data only (see text for details). \label{fig5}}
\end{figure*}

The extracted broadband SED, in the $\nu F_{\nu}$
representation, can be found in \reffig{fig5}. It 
shows a two-bump structure typical for blazars. 
Based on the location of the synchrotron peak between UV and X-rays, the
source classification as an IBL according to
\citet{2006A&A...445..441N}, or as an HSP according to
\citet{2011ApJ...743..171A}, can be confirmed. 
However, this classification might only be true for the observations
in 2011 reported here, 
as it is known that some blazars undergo spectral changes during
flares which could change their SED classification, e.g. VER~J0521+211 
\citep{verj0521}.

%
%
The SED is modeled with the synchrotron self-Compton
(SSC) model by \citet{2002ApJ...581..127B}, which assumes that the
plasma jet is 
powered by accretion of material onto a super-massive black hole
\citep[for details see][]{2009ApJ...707..612A}. 
The emission zone is modeled as a spherical volume
of radius $R$ moving with relativistic speed
$\beta_{\Gamma} c$ along the jet axis.
The jet is directed at a small angle $\theta$ with respect to the line
of sight to the observer. Since the observing angle is very hard to
measure, it is fixed within the model to the superluminal angle,
$\theta \simeq 1/\Gamma$, for which the (bulk) Lorentz factor $\Gamma$
equals the Doppler factor 
$D = (\Gamma (1-\beta_{\Gamma} \cos \theta) )^{-1}$. 
The results of the model depend mainly on the Doppler factor,
hence other combinations of $\theta$ and $\Gamma$ resulting in the
same Doppler factor are also possible.

Into the emission region, a population of ultra-relativistic
non-thermal electrons is injected 
following a power-law distribution with low- and high-energy cutoffs
$\gamma_1$ and $\gamma_2$, respectively, so that
$Q_{\mathrm{e}}(\gamma,t) = Q_{\mathrm{0}}(t) 
\gamma^{-q}$ for $\gamma_1 < \gamma < \gamma_2$. 
The normalization of the electron distribution is related to the
magnetic field $B$ through a relative partition parameter
$\epsilon_{\mathrm{B}}$, 
defined as  $\epsilon_{\mathrm{B}} = L_{\mathrm{B}}/L_{\mathrm{e}}$. 
$L_{\mathrm{e}}$ is the kinetic power in the relativistic electrons and
$L_{\mathrm{B}}$ is the power in the Poynting flux carried by the magnetic
field. The magnetic field itself is a free parameter within the
model.

As the emission region is propagating along the jet, the continuously
injected particles lose energy through synchrotron and SSC radiation 
and may escape from the emitting region. 
The particle escape is described by an escape timescale parameter 
$\eta_{\mathrm{esc}} > 1$ with $t_{\mathrm{esc}} = \eta_{\mathrm{esc}} \cdot R/c$. 
As a result of the assumed quasi-equilibrium between particle
injection, escape and radiative cooling, a break in the electron
distribution will occur self-consistently at a Lorentz factor
$\gamma_{\mathrm{b}}$, where $t_{\mathrm{esc}} = t_{\mathrm{cool}}(\gamma_{\mathrm{b}})$. 
Depending on whether $\gamma_{\mathrm{b}}$ is larger or less than $\gamma_1$, the
system will be in the slow or fast cooling regime. In the fast cooling
regime ($\gamma_{\mathrm{b}} < \gamma_1$), the equilibrium electron distribution
will be a broken power law with $n(\gamma) \propto \gamma^{-2}$ for
$\gamma_{\mathrm{b}} < \gamma < \gamma_1$ and 
$n(\gamma) \propto \gamma^{-(q+1)}$ for $\gamma_1 < \gamma < \gamma_2$. 
In the slow cooling regime ($\gamma_{\mathrm{b}} > \gamma_1$), the broken power
law is of the form $n(\gamma) \propto \gamma^{-q}$ for
$\gamma_1 < \gamma < \gamma_{\mathrm{b}}$ and 
$n(\gamma) \propto \gamma^{-(q+1)}$ for $\gamma_{\mathrm{b}} < \gamma <
\gamma_2$.

%
%
Altogether, the SSC model described here has eight free parameters,
listed in \reftab{tab:VERITASblazars}. 
Several of these parameters can be estimated from observables like the
spectra obtained from multi-wavelength (MWL) observations as well as
the measured 
variability time scales (see, e.g., \citealp{1998ApJ...509..608T}). 
The absorption of VHE gamma rays on the extragalactic background light
(EBL) is accounted for in the predicted fluxes using the EBL model  
of \citet{2010ApJ...712..238F}\footnote{This absorption is consistent
with the absorption level derived from EBL models of
\citet{2008A&A...487..837F} and \citet{2009MNRAS.399.1694G}.}. 
Here, a redshift of $z=0.130$ is used for the modeling; we 
will discuss later how the results are affected if a 
redshift of $z=0.237$ is adopted instead.

In \reffig{fig5} the results of the SSC model are shown. 
It can be seen that the overall SED for both X-ray states in 2011 can
be well described by the  model. 
The solid lines represent the model for which the archival data 
in the micrometer waveband are taken into account. 
At those frequencies of around $10^{11}$~Hz, a spectral break
occurs due to the transition from fast cooling to the
slow cooling regime. The position of this break is determined by the
escape time scale, resulting in a large value for $\eta_{\mathrm{esc}}$. 
The Doppler factor is relatively large, with $D=30$. 
Using a lower Doppler factor would
require a larger emission region and hence a lower magnetic field
strength, resulting in variability time scales 
longer than days (following causality
arguments). This lower Doppler factor scenario would contradict the
measurements of X-ray variability during the January observations. 
The magnetic field strength is found to be quite low,
resulting in a very small relative partition parameter. 
A magnetic field far below equipartition ($\epsilon_{\mathrm{B}} \approx 0.1-1$),
as found here, 
might indicate a particle-dominated jet, in which the magnetic field
in the emission region is self-generated and/or amplified in
shocks. In contrast, a magnetic field near or above equipartition
would be consistent with a Poynting-flux-dominated jet in which
magnetic field energy is transferred to particles, reaching
approximate equipartition in the high-energy emission zone.

In order to account for the two different X-ray states observed in
2011, the electron injection spectral index together with the
magnetic field strength within the modeled emission region were
changed. 
The injection index during the high X-ray state in January ($q=2.8$)
is found to be harder than during the low X-ray state in April/May
($q=3.4$).  
Under the assumption that the particles in the jet are accelerated
within relativistic shocks, the spectral change in the electron
distribution may be explained by a change in the shock field
obliquity: a larger angle between magnetic field and shock front 
results in a harder particle spectrum 
\citep[see, e.g.,][]{2009ApJ...698.1523S}. 
This change of the injection spectrum also 
leads to flux variations in the high-energy peak. 
However, neither \textit{Fermi}-LAT nor VERITAS are sensitive to those
variations given the flux level of the source during the observation
period reported here.

All of the above results are obtained by taking the
archival data at around $10^{11}$~Hz into account. 
However, due to possible variability in this waveband, we also modeled
the SED of the source ignoring those archival data. 
The resulting model spectrum is represented by the dashed lines in
\reffig{fig5} and represents the SED well. 
The only changes made to the model parameters were to reduce the value
of the escape time parameter $\eta_{\mathrm{esc}}$ and the injection power. 
It was found that a value up to ten times smaller for
$\eta_{\mathrm{esc}}$ could 
be used to model the SED. This means that the escape time scale may be
shorter and the cooling break would occur at higher energies compared to the
model including the archival data. Additionally, the system is closer to
equipartition, since the injection power for the electron distribution
is lower (with the same value for the magnetic field strength). 
Due to the lack of simultaneous data in the $10^{11}$~Hz
domain, $\eta_{\mathrm{esc}}$ and $L_{\mathrm{e}}$ are unconstrained,
resulting in a range of parameter combinations that describe the
observed SED well. It is worth noting that in this, and most other
one-zone leptonic models, the synchrotron emission from the gamma-ray
emission region is self-absorbed at millimeter and longer
wavelengths. Therefore, these models are 
often unable to account for the radio emission, which is generally
thought to be due to the superposition of self-absorbed synchrotron
components produced further out along the jet
\citep{1985ApJ...298..114M}, and is treated in one-zone emission
models as upper limit.

Another difficulty of the model and its possible interpretation is
the uncertainty in the redshift. We applied the same model to
the SED using $z=0.237$. We found that 
both X-ray states can well be modeled by a change of the electron
injection spectral index together with the magnetic field strength. 
In this case, the Doppler factor needs to be larger ($D=50$) to
compensate for the EBL absorption at high energies 
and the model-predicted fluxes are found to be slightly below the 
VERITAS spectral points. Nevertheless, this is still compatible with
the VERITAS measurement considering systematic errors, therefore, this
redshift cannot be excluded.

\section{Discussion}

As previously mentioned, B2~1215+30 was observed by MAGIC early in
2011. \citet{2012A&A...544A.142A} presented the source as 
an ``exceptional VHE $\gamma$-ray emitting BL Lac'', mainly based on
its SED and the results obtained from the MWL modeling. 
Here, the results of the modeling presented in Section~6 are compared
to those in the MAGIC publication and then put into perspective with 
results obtained from other blazars detected by VERITAS.

MAGIC observed B2~1215+30 for approximately 20~hours between January and
February 2011 and their spectral points are shown in \reffig{fig5},
compatible with the VERITAS measurements. 
Quasi-simultaneous MWL data, also compatible with those obtained here, were
used to construct the SED and were modeled by the SSC model of
\citet{2003ApJ...593..667M} using the redshift of $z=0.13$. 
In the paper, \citet{2012A&A...544A.142A} represent the data with SSC
model parameters which are compatible with those obtained here. 
They also found that the same SSC model can well reproduce the data
with a higher Doppler factor ($D=60$) or a higher minimum Lorentz
factor of the electron distribution ($\gamma_{\mathrm{min}} =
3\times10^{3}$). Using these values for the model parameters, we
did not succeed in fully representing the SED. The main reason is that
the modeled high-energy peak is not broad enough to represent the 
low-energy points of the \textit{Fermi}-LAT spectrum. However, this
part of the spectrum was not used in the MAGIC publication.

%
%
To address the question whether B2~1215+30 is extreme in terms
of its SED and model parameters, the results of our modeling are
compared to those obtained on all the VERITAS-detected blazars which
have contemporaneous MWL data and are modeled with the SSC model by
\citet{2002ApJ...581..127B}. 
In total, 6 HBLs and 3 IBLs are found; the model parameters are
listed in \reftab{tab:VERITASblazars}. Three of those blazars $-$ 
1ES~0806+524 \citep{2009ApJ...690L.126A}, Mrk~421
\citep{2009ApJ...703..169A}, and W~Comae
\citep{2008ApJ...684L..73A,2009ApJ...707..612A} $-$ were found in
different flux states during the MWL observations and have more than one
set of model parameters. 
PKS~1424+240 \citep{2010ApJ...708L.100A} and 3C~66A
\citep{2011ApJ...726...43A} have been modeled using different redshift
assumptions. While PKS~1424+240 will not be included in the comparison
study due to the current lack of redshift constraint, two redshift
values are given in \reftab{tab:VERITASblazars} for 3C~66A,
i.e. $z=0.3$ and $z=0.444$, as they enclose the recently 
published redshift limits which were found to be in the range of $0.3347 < z
\leq 0.41$ \citep{2013arXiv1302.2948F}.

As one can see in \reftab{tab:VERITASblazars}, most of the parameters
used to model the SED of B2~1215+30 are well within the range of those
used for previously detected blazars. 
The Doppler factor, for example, is usually found to be in
the range of $D = 20-30$ for the applied model. This is in agreement
with other SSC models, e.g. \citet{2010MNRAS.401.1570T}. 
However, there are two parameters which are outside this 
``standard range'': the magnetic field strength $B$ and the escape
time parameter $\eta_{\mathrm{esc}}$. The first one is relatively low for
B2~1215+30 and results in a very low relative partition parameter. For some
of the other blazars, e.g. W~Comae and 3C~66A, this behavior has also
been seen. In those cases, an SSC model with an external radiation
field resulted in model parameters with larger magnetic field
strengths and closer to equipartition. 
However, in the case of B2~1215+30 no improvement could be found by
adding an EC component to the model - neither in the representation of
the shape of the SED nor by bringing the system closer to equipartition.
In general, the magnetic field strength values obtained for the
different sources are consistent with results from other leptonic
blazar models. 
The second parameter, $\eta_{\mathrm{esc}}$, is high
compared to the model parameters of the other blazars. Such a high
value for $\eta_{\mathrm{esc}}$ implies long escape time scales. This
could hint at a relatively well ordered (laminar) magnetic field in
the emission region. 
However, it has already been shown in the previous section that the
value for $\eta_{\mathrm{esc}}$ can be lowered significantly when taking only
the contemporaneous data into account, without losing the ability to
reproduce the shape of the SED. This value is then closer
to values applied to the other VERITAS-modeled blazars. 
In summary, the model parameters derived here for the applied SSC
model are in the range of those derived from previous VERITAS
blazar modeling. In this sense B2~1215+30 is a typical VHE-detected blazar. 


\section{Summary and Conclusion}

We have presented long term observations of BL Lac B2~1215+30 at VHE energies
with VERITAS between December 2008 and May 2012. 
During these observations, the source was clearly detected and showed 
clear variability on time scales longer than months,
while variability on shorter time scales could not be detected. 
In 2011, the source was found to be in a bright state and a spectral
analysis could be performed. The results are compatible with the MAGIC
results from early 2011 reported in \citet{2012A&A...544A.142A}.

MWL data, quasi-simultaneous with the VERITAS observations in 2011, were
used to construct the SED of B2~1215+30 and confirmed its
classification as an IBL.  
During these VERITAS observations, B2~1215+30 showed different
flux states in the X-ray regime. These could be successfully
reproduced with an SSC model by changing the spectral index of the
injected electron distribution together with the magnetic field
strength. 
Our study finds a model description for the SED of B2~1215+30 similar
to other TeV-detected blazars.

Observations of B2~1215+30 by VERITAS will
continue as part of a monitoring program on 1ES~1218+304, a  
TeV blazar in the same field of view. This will allow a search
for variability on different time scales and could result in tighter
constraints for the input model parameters, as dedicated observations
of B2~1215+30 can be triggered in case of an increased flux state.

\acknowledgments
This research is supported by grants from the U.S. Department of
Energy Office of Science, the U.S. National Science Foundation and the
Smithsonian Institution, by NSERC in Canada, by Science Foundation
Ireland (SFI 10/RFP/AST2748) and by STFC in the U.K., as well as award
NNX12AJ12G from the NASA {\it Swift} Guest Investigator program. 
We acknowledge
the excellent work of the technical support staff at the Fred Lawrence
Whipple Observatory and at the collaborating institutions in the
construction and operation of the instrument. 
We are also grateful to Grant Williams and Daniel Kiminki for their
dedication to the operation and support of the Super-LOTIS telescope. 
HP acknowledges support through the Young Investigators Program of
the Helmholtz Association. 
MB acknowledges support by the South African Research Chairs
Initiative of the Department of Science and Technology and the
National Research Foundation of South Africa.
Support for MF was provided by NASA through Hubble Fellowship grant
HF-51305.01-A awarded by the Space Telescope Science Institute, which
is operated by the Association of Universities for Research in
Astronomy, Inc., for NASA, under contract NAS 5-26555.  




\clearpage

\clearpage 
\begin{deluxetable}{lccrrrrc}
\tablewidth{0pt}
\tablecaption{VERITAS results of B2~1215+30 for the three different
  observing epochs. \label{tbl-1}} 
\tablehead{
\colhead{Data set} & 
\colhead{Exposure} & 
\colhead{Zenith} & 
\colhead{N$_{\mathrm{ON}}$} &
\colhead{N$_{\mathrm{OFF}}$} & 
\colhead{$\alpha$} & 
\colhead{Sig} & 
\colhead{$F(E>200\mathrm{GeV})$ } \\
\colhead{} & 
\colhead{[hours]} & 
\colhead{[deg]} & 
\colhead{ } &
\colhead{ } & 
\colhead{ } & 
\colhead{[$\sigma$]} & 
\colhead{[$10^{-12} \mathrm{cm}^{-2} \mathrm{s}^{-1} $] }
}
\startdata
2008/09 & 29 & 20 & 304 & 2288 & 0.1243 &  1.1 & $< 4.5$ \\
2011 & 38 & 15 & 472 & 2325 & 0.1161 & 10.4 & $8.0 \pm 0.9$ \\
2012 & 15 & 12 & 143 &  898 & 0.1177 &  3.2 & $2.8 \pm 1.1$ \\
\tableline
TOTAL & 82 & 16 & 919 & 5511 & 0.1197 & 8.9 & $-$ \\
\enddata
\tablecomments{The exposure time in hours is given in effective time
  on B2~1215+30 (in 0.5\degr\ wobble offset equivalent). 
  Zenith refers to the mean zenith angle of the observations. 
  N$_{\mathrm{ON}}$ and N$_{\mathrm{OFF}}$ are the number of events in the
  ON and OFF region, while $\alpha$ is the acceptance-corrected area
  ratio of both regions. 
  Sig is the significance of the detection of B2~1215+30. 
  The flux (or 99\% upper limit) above 200~GeV is calculated with
  $1\sigma$ statistical error bars using the spectral index derived in 2011.
}
\end{deluxetable}

\clearpage

\begin{deluxetable}{lrrrrrrrrr rr r}
\tabletypesize{\scriptsize}
\tablewidth{0pt}
\tablecaption{SSC model parameters for B2~1215+30 for the high (H) and
  low (L) X-ray states in 2011 (as
  explained in the text) and other VERITAS-detected blazars using
  the same model. \label{tab:VERITASblazars} }
\tablehead{
\colhead{Source} & 
\colhead{$z$} & 
\colhead{$L_e $} & 
\colhead{$\gamma_1$} & 
\colhead{$\gamma_2$} & 
\colhead{$q$} & 
\colhead{$B $} & 
\colhead{$R $} & 
\colhead{$D$} & 
\colhead{$\eta_{\mathrm{esc}}$} & 
\colhead{$\epsilon_{\mathrm{B}}$} & 
\colhead{$\delta t_{\mathrm{var}}$} &
\colhead{ref} \\

\colhead{} & \colhead{} & 
\colhead{[$10^{44}$~erg/s]} & 
\colhead{[$10^{3}$]} & \colhead{[$10^{5}$]} & 
\colhead{} & 
\colhead{[G]} & 
\colhead{[$10^{16}$~cm] } & 
\colhead{} & \colhead{} & \colhead{} & 
\colhead{[hr]} &
\colhead{}}

\startdata

\textbf{B2 1215+30} (H) &
0.13\tablenotemark{*} &
5.0\tablenotemark{a} &
30 &
8 &
2.8 &
0.02 &
5 &
30 &
3300\tablenotemark{a} &
0.00678\tablenotemark{a} &
17.4 &\\

\textbf{B2 1215+30} (L) &
0.13\tablenotemark{*} &
6.5\tablenotemark{b} &
45 &
20 &
3.4 &
0.01 &
13 &
30 &
2500\tablenotemark{b} &
0.00874\tablenotemark{b} &
45.3 &\\

\tableline

RBS~0413 & 0.19 & 0.297 & 70 & 10 & 2.4 & 0.1 & 1.1 & 20 & 10 & 0.06 &
6.06 & 1\\
1ES~0414+009 & 0.287 & 3.07 & 200 & 50 & 3.5 & 0.008 & 21 & 40 & 120 &
0.055 & 62.5 & 2\\
RX~J0648+1516 & 0.179 & 0.75 & 67 & 10 & 4.8 & 0.14 & 2 & 20 & 100 &
0.16 & 10.9 & 3\\ 
RGB~0710+591 & 0.125 & 0.449 & 60 & 20 & 1.5 & 0.036 & 2 & 30 & 100 &
0.039 & 6.9 & 4\\
1ES~0806+524 & 0.138 & 0.19 & 17.7 & 2 & 3.1 & 0.39 & 0.5 & 20 & 3 &
0.31 & 2.6 & 5\\
1ES~0806+524 & 0.138 & 0.14 & 16 & 2 & 2.7 & 0.5 & 0.5 & 20 & 3 & 0.68
& 2.6 & 5\\
Mrk 421 & 0.031 & 0.0776 & 42 & 5 & 2.6 & 0.48 & 0.3 & 20 & 3 & 0.4 & 1.4 &6\\
Mrk 421 & 0.031 & 0.106 & 33 & 4 & 3.2 & 0.68 & 0.3 & 20 & 3 & 0.59 & 1.4 &6\\


W Comae & 0.102 & 28 & 0.45 & 4.5 & 2.2 & 0.007 & 10 & 30 & 200 &
0.00059 & 31.2 &7\\ 

W Comae & 0.102 & 3.4 & 9 & 2.5 & 2.55 & 0.24 & 0.3 & 20 & 300 &
0.0023 & 1.5 &8\\

3C 66A & 0.3\tablenotemark{*} & 57 & 22 & 4 & 3 & 0.02 & 7 & 40 & 1000
& 0.0015 & 21.1 & 9\\ 
3C 66A & 0.444\tablenotemark{*} & 128 & 25 & 5 & 3 & 0.01 & 11 & 50 &
1000 & 0.0011 & 29.4 &9\\


\enddata

\tablenotetext{a}{Alternative parameters without archival data are $L_{\mathrm{e}}=2.9$, $\eta_{\mathrm{esc}}=330$, and $\epsilon_{\mathrm{B}}=0.0116$.}
\tablenotetext{b}{Alternative parameters without archival data are $L_{\mathrm{e}}=3.6$, $\eta_{\mathrm{esc}}=250$, and $\epsilon_{\mathrm{B}}=0.0159$.}

\tablecomments{The columns are the following: $z$ is the redshift 
(* denotes the assumed redshift within the model, see text for more
details); 
$L_{\mathrm{e}}$ is the jet luminosity; $\gamma_1$ is the low
energy cutoff energy of the electron distribution; $\gamma_2$ is the
high energy cutoff;
$q$ is the spectral index of the electron injection spectrum; $B$ is
the magnetic field strength; $R$ is the emission region radius; $D$ is
the Doppler factor; and 
$\eta_{\mathrm{esc}}$ is the escape time parameter. Additionally, two output
parameters are given: $\epsilon_{\mathrm{B}}$ is the
resulting relative partition parameter $\epsilon_{\mathrm{B}} =
L_{\mathrm{B}}/L_{\mathrm{e}}$; and $\delta t_{\mathrm{var}}$ is the
resulting minimum variability time scale. } 

\tablerefs{
(1) \cite{2012ApJ...750...94A}; 
(2) \cite{2012ApJ...755..118A};
(3) \cite{2011ApJ...742..127A};
(4) \cite{2010ApJ...715L..49A};
(5) \cite{2009ApJ...690L.126A};
(6) \cite{2009ApJ...703..169A};
(7) \cite{2008ApJ...684L..73A};
(8) \cite{2009ApJ...707..612A};
(9) \cite{2011ApJ...726...43A};
}
\end{deluxetable}

\end{document}